# Determination of Edge Purity in Bilayer Graphene Using μ-Raman Spectroscopy

Milan Begliarbekov[1], Onejae Sul[2], Sokratis Kalliakos[1], Eui-Hyeok Yang[2], Stefan Strauf[1]*

[1]Department of Physics & Engineering Physics, Stevens Institute of Technology, Hoboken NJ, USA and
[2]Department of Mechanical Engineering, Stevens Institute of Technology, Hoboken NJ, USA

Polarization resolved μ-Raman spectroscopy was carried out at the edges of bilayer graphene. We find strong dependence of the intensity of the G band on the incident laser polarization, with its intensity dependence being 90° out of phase for the armchair and zigzag case, in accordance with theoretical predictions. For the case of mixed-state edges we demonstrate that the polarization contrast reflects the fractional composition of armchair and zigzag edges, providing a monitor of edge purity, which is an important parameter for the development of efficient nanoelectronic devices.

The recent discovery of graphene [1], a two-dimensional crystal comprised of a single layer of carbon atoms, triggered intensive research efforts in the physics and materials science communities. The high degree of crystallinity and outstanding electronic and thermal properties make graphene a promising candidate for nanoelectronic devices [2–4]. The addition of a second layer forms bilayer graphene with a largely changed electronic band structure resulting in field-tunable electronic band gaps [5] and strongly suppressed electronic noise [6]. Of particular importance for device applications are the underlying edge chiralities of bilayer graphene and graphene nanoribbons (GNRs), since the atomic edge composition influences the electronic structure and thus transport properties [7, 8] as well as chemical reactivity [9]. As a nondestructive technique, Raman spectroscopy has been widely utilized to determine the number of graphitic layers [10, 11]. Furthermore, since the chirality of graphitic edges and the orientation of the crystalline axis have a strong impact on phonon modes localized at the edges, Raman spectroscopy can also be utilized for edge state characterization [12–15]. Although previous experiments have addressed the issue of edge state identification by Raman spectroscopy using the D band around 1350 $cm^{-1}$ [13], a detailed analysis and methodology to determine edge purity in the case of mixed edges has not yet been presented. Unlike the D band, the G band around 1580 $cm^{-1}$ was recently predicted to show a strong polarization sensitivity with respect to armchair and zigzag edges, with Raman scattering amplitudes 90 degrees out of phase [15].

Here, we report on polarization-resolved μ-Raman experiments performed at the edges of bilayer graphene flakes. We find a strong dependence of the Raman intensity of the G-band on the polarization of incident laser light with respect to various edge orientations and we confirm that amplitudes of armchair and zigzag edges are 90° out of phase. Furthermore, we demonstrate that the varying polarization contrast of the G band is a useful monitor to characterize edges with mixed armchair/zigzag boundaries.

In these experiments, graphene flakes were mechanically exfoliated from a highly ordered pyrolyzed graphite (HOPG) block and deposited onto pre-patterned p$^{++}$ silicon wafer with a thermally grown 300 nm silicon oxide. Room temperature μ−Raman spectra were obtained using a 2.33 eV laser diode with a spot size of about 2 μm. Half wave plates were used to rotate the plane of polarization with respect to the sample in the laser excitation path and to rotate the plane of polarization in the collection path back to its original configuration in order to eliminate any errors introduced by the dependence of the spectrometer's grating and other optical components on the polarization of light.

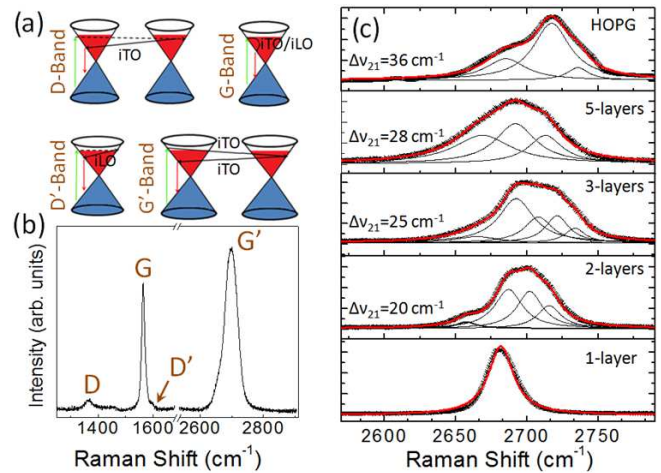

Figure 1: a) The scattering mechanisms that give rise to the various Raman modes, here solid horizontal lines show phonon scattering, whereas dashed horizontal lines show defect scattering; b) abridged Raman spectrum labeling the bands identified in a; c) G' band Raman spectra obtained from several different flakes (offset for clarity) showing the dependence of the G' band on the number of graphitic layers. The solid red line is the sum of the Lorentzian sub-components.

The prominent spectral bands of graphene are shown in the Raman spectrum in Fig. 1b., while Fig. 1a shows the physical mechanisms that give rise to these bands. Each band can be used as a tool to probe different material characteristics. The G' band (sometimes referred to as the 2D band) provides unambiguous information about the number of constituent graphene layers. This phonon band (2700 $cm^{-1}$) originates from inter-valley scattering of two in-plane transverse optical (iTO) phonons at the K and K' points at the edges of the Brillouin zone [16, 17]. The impact of the number of layers on the G' band is shown in Fig 1c. In single-layer graphene, the G' band can be approximated by a single Lorentzian function (Fig. 1c, lower panel), whereas several Lorentzian functions are required in the case of multilayer graphene (Fig.

*Electronic address: strauf@stevens.edu

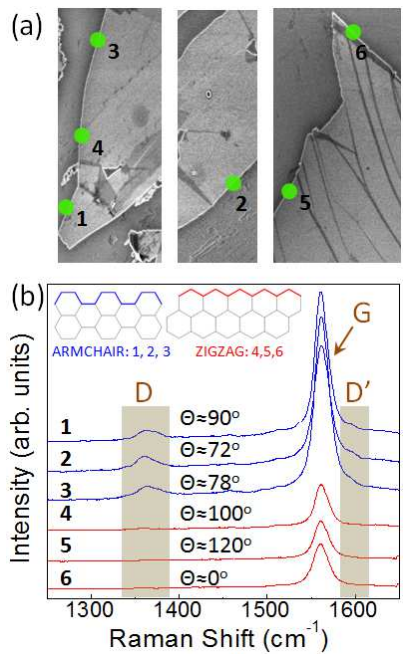

Figure 2: a) Scanning electron micrographs of different regions of the graphene flake from which the spectra were obtained. b) Raman spectra of the edges identified in 'a,' (offset for clarity). The angle Θ is measured between edge 6 and the edge from which the spectra were obtained. The presence (absence) of the D and D' bands is indicative of armchair (zigzag) edge chiralities.

1c, upper panels), reflecting the splitting of the electronic bands and phonon branches [18]. After peak deconvolution we find that the difference in frequencies of the two dominant subcomponents of the G' band $\Delta\nu_{21}$ increases with the number of graphitic layers, with values comparable to the ones reported in the literature [10]. The following investigation focuses on exfoliated flakes which have been identified as bilayer graphene.

While the G' band is useful in layer metrology analysis, the D and D' bands can be used for edge chirality determination. Figure 2b shows Raman spectra of different edges of a bilayer flake obtained under the same polarization conditions. All edges were selected from a single large area flake, as shown in Fig. 2a. Edge 6 was identified as being zigzag using D / D' band spectroscopy (see discussion below) and used as a reference for measuring all subsequent edge angles, identified as Θ in Fig. 2. Interestingly, several of the edges possess pronounced D (1350 $cm^{-1}$) and D' (1620 $cm^{-1}$) bands, while others lack both bands. The D band originates from intervalley scattering that connects two adjacent K & K' points at the Brillouin zone boundary via a second order process that requires one iTO phonon and a symmetry breaking perturbation such as an armchair edge for its activation [17, 19]. Similarly, the D' band is a weak intra-valley transition that requires one iLO phonon and a symmetry break [16, 17]. The presence (absence) of these bands has been shown to correspond to armchair (zigzag) chiralities in both single layer graphene [13] as well as HOPG [20]. Based on the fact that the chirality of a given edge changes in multiples of 30° (with odd multiples corresponding to edges with opposite chirality and even multiples corresponding to edges with the same chirality [13]), we identify edge 5, being 120°degrees with respect to edge 6 (a zigzag edge) as zigzag, while edge 1, which is 90° with respect to edge 6 as an armchair edge. Other edges can be identified as either predominantly zigzag or armchair depending if the angle they make with respect to the edge is closer to an even or an odd multiple of 30°. Thus the angle metrology and the correlation with the presence or absence of the D-band allows us to make a distinction between armchair and zigzag edges.

However, the D band does not provide unambiguous information about edge purity. For example, a lower purity is expected for the case of edge 2 and 3 with 72° and 78° respectively as is evident from the schematic in Fig. 3 c, but the D band does not change its oscillator strength accordingly and was found not to exhibit strong polarization dependence.

Following the initial identification of the edge chiralities in our sample, we now focus on the G band around 1580 $cm^{-1}$. The G band arises from a doubly degenerate intra-valley process that originates from scattering of an iTO phonon or an iLO phonon at the center (Γ-point) of the Brillouin zone [16, 17]. For pure zigzag edges, the intensity of the G band is expected to be maximum for an excitation beam polarization that is perpendicular to the edge. Conversely, for armchair edges its intensity maximizes for the incident excitation beam polarization that is parallel to the edge. This phenomenon is still present for mixed edges, however, the degree of the polarization contrast is diminished and is proportional to the amount of mixing of zigzag and armchair boundaries. Purely random edges, i.e. edges comprised of equal amounts of zigzag and armchair boundaries, are not expected to exhibit any polarization dependence [15].

The polarization dependence of the G band obtained from edges 1 (armchair) and 6 (zigzag) is shown in Fig. 3a. We find that the intensity of the G band of the armchair and zigzag edges has a strong polarization dependence, that is 90°out of phase with respect to each other. The intensity dependence of the armchair edge varies according to $I_a^G \propto \sin^2 \varphi$ (solid red line in Fig. 3a), while for the zigzag edge it varies according to $I_z^G \propto \cos^2 \varphi$ (solid blue line in Fig 3a), where $\varphi$ is the angle between the edge of the flake and the polarization axis of the excitation beam [15]. Note that the data in Fig. 3a have been corrected for a nonvanishing background of about 2800 counts to emphasize the polarization contrast, while Fig. 3b shows raw data without any background substraction.

Furthermore, the G band shows no polarization dependence far from the edges ($\geq 3\mu$m), as shown by the black circles, obtained at the center of the flake. Similar nonpolarized data were obtained at numerous different points away from the edges and across the entire flake. Earlier experiments on the G band found a variation in amplitude when scanning across a flake at various interior points [12, 21], which is related to Kohn anomalies and an underlying non-uniform strain or deformation potential [22]. Consequently, the lack of polarization dependence at interior points (in basal plane of graphene) is indicative that the observed phenomenon in our experiments arises from the different allowed and forbidden phonon modes at the edges of the flakes and not from strain-related effects.

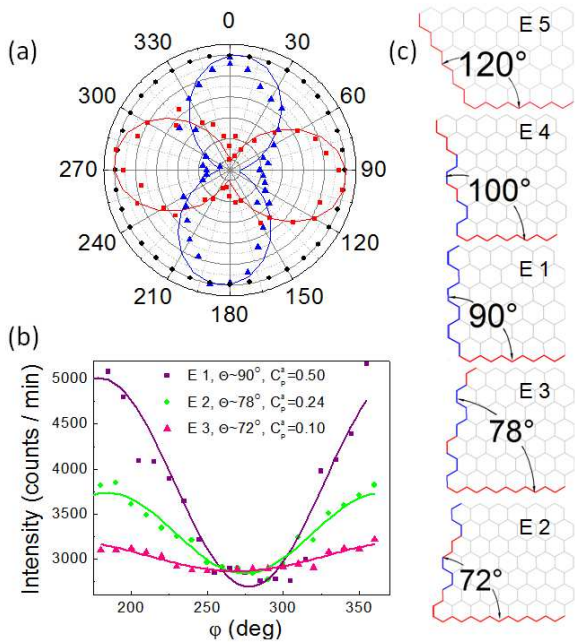

Figure 3: a) Polarization dependence of the edges E1 (armchair, blue triangles) and E6 (zigzag, red squares). The black circles correspond to the data obtained at the center of the flake, which shows no polarization contrast. b) polarization contrast of 3 armchair edges (E1, E2, & E3) showing variable polarization contrast $C_p^a$ which is correlated to the expected amount of zigzag contamination at that edge; c) schematic representation of the edges

The presence (absence) of the D band is strongly correlated to 30° multiplicity of the edges as shown above. This effect originates from the fact that only the longitudinal (transverse) optical phonon mode is a Raman active mode near the armchair (zigzag) edge. Since the physical mechanism that gives rise to the G band originates from scattering of a doubly degenerate iTO and an LO phonon at the Brillouin zone boundary, the G band should be better suited for the mapping of edge states with different or mixed chirality. To this end we recorded the polarization contrast of the G band in 3 different armchair edges (Fig. 3b). We find strong correlation of the relative intensity change with the multiplicity of those edges. In contrast, the intensity of D band showed little sensitivity on the incident photon polarization (data not shown). More precisely, edge 1, which is 90° to the dominant zigzag edge has the highest (50 %) polarization contrast and the closest odd multiplicity $\left(\frac{90}{30} = 3\right)$ while edges 2 and 3 have 26% and 10% polarization contrasts with multiplicities of $\frac{78}{30} = 2.6$ and $\frac{72}{30} = 2.4$ respectively. It should be noted that, in all cases, the polarization dependence of the G-band tends to a minimum value, but never vanishes, suggesting that although the edge is comprised of mostly armchair constituents, it is not atomically clean within the detection area (2 $\mu$m spot size). This verifies prior experimental results [13], which show that atomically smooth edges are very rarely obtained using micromechanical exfoliation.

In summary, we found that the Raman G band in bilayer graphene is particularly sensitive to the laser polarization with its intensity dependence being out of phase by 90° in the armchair and zigzag case. In addition, for mixed-state edges we observe that the G band polarization contrast reflects the fractional composition of armchair and zigzag edges and provides thus information about the purity of the edge. This knowledge is crucial for the development of graphene-based electronic devices and could serve as a convenient process monitor to characterize the degree of edge state purity in GNRs created with various fabrication techniques such as exfoliation, electron beam lithography, or local anodic oxidation.

During the review process of this manuscript we became aware of a recent work by Cong et al. [23], showing a similar polarization dependence of the G-band using monolayer graphene. The combined knowledge of our work and the work by Cong et al. suggest that the polarization dependence in mono- and bilayer graphene is of the same origin. In addition, our work considers the case of mixed edges which are most relevant for technological applications.

### Acknowledgments

We would like to thank Steve Tsai, Johanna Heureaux, and Anderson Tsai for their aid with sample preparation. Partial financial support for this work was provided by the NSF GK-12 Grant No. NSF DGE-0742462 and by the Air Force Office for Scientific Research (award no FA9550-08-1-0134).

[1] K. Novoselov, A. Geim, S. Morozov, D. Jiang, Y. Zhang, et al., Science **306**, 666 (2004).
[2] K. I. Bolotin, K. J. Sikes, J. Hone, H. L. Stormer, and P. Kim, Phys. Rev. Lett. **101**, 096802 (2008).
[3] A. C. Neto, F. Guinea, N. Peres, K. Novoselov, and A. Geim, Rev. Mod. Phys **81**, 109 (2009).
[4] K. S. Novoselov, D. Jiang, F. Schedin, T. J. Booth, V. V. Khotkevich, et al., Proc. Nat. Acad. Sci. **102**, 10451 (2005).
[5] Y. Zhang, T.-T. Tang, C. Girit, Z. Hao, M. Martin, et al., Nature **459**, 820 (2009).
[6] Y. Lin and P. Avouris, Nano Lett. **8**, 2119 (2008).
[7] Y. Miyamoto, K. Nakada, and M. Fujita, Phys. Rev. B **59**, 9858 (1999).
[8] K. A. Ritter and J. W. Lyding, Nature Mat. **8**, 235 (2009).
[9] C. O. Girit, J. C. Meyer, R. Erni, M. D. Rossell, C. Kisielowski, et al., Science **323**, 1705 (2009).
[10] D. Graf, F. Molitor, K. Ensslin, C. Stampfer, A. Jungen, et al., Nano Lett. **7**, 238 (2007).
[11] A. C. Ferrari, J. C. Meyer, V. Scardaci, C. Casiraghi, M. Lazzeri, et al., Phys. Rev. Lett. **97**, 187401 (2006).
[12] K. Sasaki, M. Yamamoto, S. Murakami, R. Saito, M. S. Dresselhaus, et al., Phys. Rev. B **80**, 155450 (2009).
[13] Y. M. You, Z. H. Ni, T. Yu, and Z. X. Shen, Appl. Phys. Lett. **93**, 163112 (2008).
[14] C. Casiraghi, A. Hartschuh, H. Qian, S. Piscanec, C. Georgi, et al., Nano Lett. **9**, 1433 (2009).
[15] K. Sasaki, R. Saito, K. Wakabayashi, and T. Enoki, J. Phys. Soc. Jpn. **79**, 044603 (2010).

[16] L. M. Malard, M. A. Pimenta, G. Dresselhaus, and M. S. Dresselhaus, Phys. Rep. **47**, 51 (2009).
[17] D. M. Basko, S. Piscanec, and A. C. Ferrari, Phys. Rev. B **80**, 165413 (2009).
[18] D. L. Mafra, L. M. Malard, S. K. Doorn, H. Htoon, J. Nilsson, et al., Phys. Rev. B **80**, 241414(R) (2009).
[19] A. C. Ferrari and J. Robertson, Phys. Rev. B **61**, 14095 (2000).
[20] L. G. Cancado, M. A. Pimenta, B. R. A. Neves, M. S. S. Dantas, and A. Jorio, Phys. Rev. Lett. **93**, 247401 (2004).
[21] J. Yan, E. A. Henriksen, P. Kim, and A. Pinczuk, Phys. Rev. Lett. **101**, 136804 (2008).
[22] M. Huang, H. Yan, C. Chen, D. Song, T. F. Heinz, et al., Proc. Nat. Acad. Sci. **106**, 7304 (2009).
[23] C. Cong, T. Yu, and H. Wang, ACS Nano, in press, DOI: 10.1021/nn100705n, (2010).